\documentclass[aps,pre,twocolumn,groupedaddress,showpacs,psfig]{revtex4}
\usepackage{amsmath}
\usepackage{psfig} 
 
\begin{document}

\title{Bridge-mediated Donor-Acceptor Effective Coupling: Exact theoretical description.}

\author{\sc D. J. Bicout$^{\ast , 1 , 2}$, F. Varchon$^{1}$, and E. Kats$^{1 , 3}$}

\affiliation{$^1$Institut Laue-Langevin, 6 rue Jules Horowitz,
B.P. 156, 38042 Grenoble, France \\
$^{2}$Biomathematics and Epidemiology, ENVL, B.P. 83,
69280 Marcy l'Etoile, France \\
$^3$
L. D. Landau Institute for Theoretical Physics,
RAS, 117940 GSP-1, Moscow, Russia.}

\begin{abstract} 
An exact expression of the bridge-mediated donor-acceptor effective coupling, 
$H_{da}$, is derived. For systems represented by a tight-binding Hamiltonian 
with nearest-neighbor interactions, we show that $|H_{da}|^2$ is equal to the 
product over all square of nearest-neighbor couplings divided by an appropriate 
product of level spacing of eigen energies of the Hamiltonian. 
Results of this calculation are compared to those obtained by perturbative approaches
and some drawbacks of the latter are pointed out.
\end{abstract}

\pacs{34.70.+e , 73.40.Gk , 72.20Ee}
 
\date{\today}

\maketitle 

\renewcommand{\theequation}{\arabic{equation}} 
\newcounter{xeq}

\setcounter{equation}{0}

There are many physical, chemical and biological processes involving transfers  
of excitations between electrodes or from a donor (D) to an acceptor (A), through a mediated 
(environmental) composite bridge (B), that can be described formally by the 
reaction scheme, $D^{\star}-B-A\:\longrightarrow\:D-B-A^{\star}$. Such processes 
include, for instance, charge transfer, where charge are transfered 
from the donor to the acceptor through intermediate states connected them, 
or in energy transfer \cite{MA56,McC61,KPU78}.  
The quantity of interest for such donor-bridge-acceptor systems is the effective 
coupling $H_{da}$ between the donor and the acceptor
which carries the dominant distance dependence of charge transfer efficency,
and that enters the transfer 
rate constant as a principal factor among others. Although the problem of 
computing the effective coupling is general to many processes, to be specific we will present 
our calculations within the framework of charge transfer. 
Recently, a number of experimental groups have reported measurements of the
DNA charge transfer, and besides several theoretical
models have been developed (see e.g., \cite{HR02}, \cite{JR97}, more
recent review papers \cite{EC04}, \cite{PC04}, and references herein).
It is however disappointing that the results obtained so far using different approaches
are not the same and even irreconcilable.
It would therefore seem appropriate at this time
to put the theoretical treatment on a firm footing to analyze more accurately all essential ingredients of the
charge transfer phenomena, and among those the
principal one - the effective $D$ - $A$ coupling $H_{da}$ is our concern in this paper.
In what follows we derive an exact close analytical expression for $H_{da}$. 
It is not only of intellectual interest but also of relevance to gain further insight
into the nature of DNA charge transfer, since comparing our exact expression and known in the literature results,
one can estimate the accuracy and reliability of various theoretical approximations. 

For a two-state system where the system is initially prepared in the donor state, 
the probability of finding the system in the acceptor state is given by,
\begin{equation}
P_{d\rightarrow a}(t)=\sin^2\left[\frac{(E_+-E_-)t}{2\hbar}\right]\:,
\label{pda}
\end{equation}
where $E_{\pm}$ (with $E_+>E_-$) are the splitting eigen energies of the 
Hamiltonian describing the system. Using the relation, 
$\displaystyle\lim_{t\rightarrow\infty}\sin^2(xt)/x^2=\pi t\delta(x)$, yields, 
$\displaystyle\lim_{t\rightarrow\infty}P_{d\rightarrow a}(t)\simeq k_{da}t$ 
where $k_{da}$ is the rate constant given by the Golden rule, 
$k_{da}=\frac{2\pi}{\hbar}\,|H_{da}|^2\,\delta(E_+-E_-)$, where 
$H_{da}=(E_+-E_-)/2$ is the effective donor-acceptor coupling of the problem. 
For multi-state systems, several authors 
\cite{McC61,LO62,LA81,DG85,EK92,MKR94,EK93,OK98}  have used various approaches 
(perturbation theory, L\"owdin's matrix partitioning technique, Green's function method) 
to derive an expression for the effective coupling. These approaches consist 
of mapping the multi-state eigenvalue problem onto the eigenvalue problem 
for the two-state system in order to determine the approximate splitting 
energies $E_{\pm}$ and $H_{da}$ as above. Accordingly, it is well established 
that the effective coupling between the donor and the acceptor can be expressed 
in terms of the Green's function of the bridge evaluated for the donor/acceptor 
energy \cite{EK92,MKR94}. 

To move further on smoothly and without loss of generality, we consider a $N+2$ 
system, $\{|n\!>\}$ denoting the localized states on the the donor ($n=0$), 
bridge ($N$ sites with one state per site, denoted $n=1,2,\cdots,N$) and acceptor ($n=N+1$). As pioneered by 
McConnell \cite{McC61}, we deal with the tight-binding (H\"{u}ckel) Hamiltonian describing 
the charge transfer from a donor to an acceptor through the bridge with site 
energies $\varepsilon_{n}$ and nearest-neighbor interactions $v_{n}$ between sites, 
\begin{eqnarray}
H=\left(\begin{array}{ccccc}
\varepsilon_0 & v_1 & 0 & 0 & \cdots \\
v_1 & \varepsilon_1 & v_2 & 0 & \cdots\\
0 & \ddots & \ddots & \ddots & 0\\
\cdots & 0 & v_{N} & \varepsilon_{N} & v_{N+1} \\
\cdots & 0  & 0 & v_{N+1} & \varepsilon_{N+1} \\
\end{array}\right)\:.
\label{Ham} 
\end{eqnarray}
For this Hamiltonian (in which $\varepsilon_0=\varepsilon_{N+1}=\varepsilon$), 
the splitting energies $E_{\pm}$ can be obtained as roots of the equations \cite{DG85}, 
\begin{eqnarray}
E_{\pm}-\varepsilon=\frac{-a(E_{\pm})\pm\sqrt{[b(E_{\pm})]^2+4c(E_{\pm})}}{2}
\label{roots}
\end{eqnarray}
where, $a(E)=v_1^2G_{11}(E)+v_{N+1}^2G_{NN}(E)$, 
$b(E)=v_1^2G_{11}(E)-v_{N+1}^2G_{NN}(E)$, 
$c(E)=v_1^2G_{11}(E)+v_{N+1}^2G_{NN}(E)$, and 
$G_{nm}(E)=<\!n|(H_0-E)^{-1}|m\!>$ is the 
element of the  Green's function restricted to the bridge described by its 
Hamiltonian $H_0$. Unfortunately, the solutions to these equations are too 
cumbersome. Thus, two main approximations are done in order to derive a useful 
expression of the effective coupling. First, for not too small bridge, it is
tempting to assume 
the following inequalities between the matrix elements of the Green's function, 
$G_{11}(E_{\pm})\gg G_{1N}(E_{\pm})$ and $G_{NN}(E_{\pm})\gg G_{1N}(E_{\pm})$.
If these assumptions are granted, neglecting the term $c(E)$ in Eq.(\ref{roots}) leads to define an auxiliary 
energy $E_0$ such that, $E_0=\varepsilon-v_1^2G_{11}(E_0)
=\varepsilon-v_{N+1}^2G_{NN}(E_0)$ (or equivalently, 
$v_1^2G_{11}(E_0)=v_{N+1}^2G_{NN}(E_0)$). Now, introducing this 
$E_0$ back into Eq.(\ref{roots}) gives, $E_{\pm}\approx E_0\pm v_1G_{1N}(E_0)v_{N+1}$. 
Second, as it is still rather complicated to obtain $E_0$, an additional 
approximation consists in setting $E_0\approx \varepsilon$. Now, by the virtue of 
these two approximations, the effective coupling simply reads, 
\begin{equation}
|H_{da}^{(1)}|=\left|\frac{E_+-E_-}{2}\right|=\left|v_1G_{1N}(\varepsilon)v_{N+1}\right|
\label{hda1}
\end{equation}
This is the well known formula, and the same result was arrived
at by numerous methods \cite{EK92,MKR94}, \cite{RA90}. The superscript ``$(1)$'' refer to 
a sort of first order perturbation. It is so useful in practice that it 
is often used without care of whether the above associated approximations are 
still valid or not. For instance, Eq.(\ref{hda1}) diverges at the resonance
in the limit where the 
donor/acceptor energy coincides with eigen energies of the bridge. As shown by 
Zhdanov, in general Eq.(\ref{hda1}) may overestimate the exact value by several 
orders of magnitude \cite{ZH02}. Unfortunately, only a particular
model for the electron potential energy has been investigated in \cite{ZH02},
and this author did not provide a general 
exact expression for the effective coupling. 

In this paper, we show that an exact expression for the effective coupling that 
does not suffer of above restrictions can be worked out by using the long time 
limit of the time dependent solution of the Schr\"odinger equation along with 
the Golden rule. To proceed, we solve the time dependent problem to determine 
the probability of charge transfer defined as, 
$P_{d\rightarrow a}(t)=|<\!0|\Psi(t)\!>|^2$, where $|\Psi(t)\!>$ is the solution 
of the Schr\"odinger equation,
$$
i\hbar\,\frac{d|\Psi(t)\!>}{d t} = H|\Psi(t)\!>\, 
$$
with the initial condition, $|\Psi(0)\!>=|0\!>$. If we denote by $|\phi_n\!>$ 
and $E_n$ (with, $n=0,1,\cdots, N+1$) the eigenstates and associated eigen energies 
of a given $(N+2)\times(N+2)$ Hamiltonian, one can show that the transition 
probability reads as,   
\begin{eqnarray}
& & P_{d\rightarrow a}(t)=-\sum_{n=0}^{N}\sum_{m=n+1}^{N+1}
4<\!N+1|\Pi_n|0\!>\nonumber \\
& & \times\,<\!N+1|\Pi_m|0\!>\sin^2\left[\frac{(\Delta E_{n,m})t}{2\hbar}\right]\:,
\label{prob}
\end{eqnarray}
where, $\Delta E_{n,m}=E_n-E_m$, is the eigen energy spacing and,  
$\Pi_n=|\phi_n\!><\!\phi_n|$, is the projection operator on the eigenstate 
$|\phi_n\!>$. Equation (\ref{prob}) is a generalization of Eq.(\ref{pda}) to many 
state systems. As $t\rightarrow\infty$, the leading term in $P_{d\rightarrow a}(t)$ 
will be associated to the smallest frequency $|\Delta E_{n,m}|/2\hbar$ corresponding 
to the smallest energy spacing, $|\Delta E_{l+1,l}|$, and defining hence the index $l$. 
Likewise, proceeding as done below Eq.(\ref{pda}), we find for Eq.(\ref{prob}) that, 
$\displaystyle\lim_{t\rightarrow\infty}P_{d\rightarrow a}(t)\simeq k_{da}t$, with 
the rate constant, 
$k_{da}=\frac{2\pi}{\hbar}\,|H_{da}|^2\,\delta(E_{l+1}-E_l)$, where the effective 
donor-acceptor coupling is given by,
\begin{eqnarray}
|H_{da}|^2 & = & -<\!N+1|\Pi_{l}|0\!>\nonumber \\
& & \times\,<\!N+1|\Pi_{l+1}|0\!>(\Delta E_{l,l+1})^2\:.
\label{hda2}
\end{eqnarray}
This expression provides a quite general and close formula for computing 
the effective coupling in various contexts. 

Our main result can be formulated as follows. For a given Hamiltonian, compute the 
eigen energy spectrum $E_n$, sort the $E_n$ in ascending (or descending) order, and 
identify the index ``$l$'' defined such that 
$|\Delta E_{l+1,l}|=\mbox{min}\left[|\Delta E_{n,m}|\:;n,m\in\mbox{spectrum}\right]$. 
Next, determine the associated eigenstates $|\phi_l\!>$ and $|\phi_{l+1}\!>$, and 
finally compute the effective coupling according to Eq.(\ref{hda2}). In addition, 
an approximate expression of Eq.(\ref{hda2}) can also be derived by using the 
perturbation theory or Green's function technique for the projection operators  
$\Pi_{l}$ and eigen energies $E_l$. We expect that the $|H_{da}|$ derived this way 
to be a better approximation because of the smallest eigen energy spacing 
$\Delta E_{l,l+1}$. 

Now, to illustrate the results in Eqs.(\ref{prob}) and (\ref{hda2}), we derive 
explicitly the above expression for a tight-binding Hamiltonian as given 
in Eq.(\ref{Ham}). To this end, we consider the wave function of the form, 
$|\Psi(t)\!>=\sum_{n=0}^{N+1}c_n(t)|n\!>$, where $c_n(t)$ are the time dependent 
amplitude of the probability of charge at the $n$th site. This yields the 
following system of equations,
\begin{eqnarray} 
\left\{\begin{array}{ccl}
i\hbar\,\displaystyle{\frac{dc_0}{dt}} & = & \varepsilon_0 c_0+v_1 c_1\\ \\
i\hbar\,\displaystyle{\frac{dc_n}{dt}} & = & v_n c_{n-1}+\varepsilon_n c_n+v_{n+1}c_{n+1}
\:;\:1\leq n\leq N\\ \\
i\hbar\,\displaystyle{\frac{dc_{N+1}}{dt}} & = & v_{N+1}c_N+\varepsilon_{N+1}c_{N+1}
\end{array}\right.
\label{dN} 
\end{eqnarray}
with the initial condition, $c_n(0)=\delta_{n0}$. We define the Laplace 
transform $\widehat{f}(s)=\int_{0}^{\infty}\!dt\,f(t)\,
{\rm e}^{-st}$ of any function $f(t)$. Laplace transforming 
Eq.(\ref{dN}) and solving the resulting equation, we obtain the recurrence 
formula, 
\begin{equation}
\frac{iv_{N+1}}{\hbar}\,\Delta_{n-1}\widehat{c}_{n+1}(s)+
\Delta_n\widehat{c}_n(s)=\prod_{m=1}^{n}\left(-\frac{iv_m}{\hbar}\right)\:,
\label{rec}
\end{equation}
where $\Delta_n=(s+i\varepsilon_n/\hbar)\Delta_{n-1}-(iv_n/\hbar)^2\Delta_{n-2}$ 
with $\Delta_{-1}=1$ and $\Delta_0=(s+i\varepsilon_0/\hbar)$. 
If$E_n$ $(n=0,1,\cdots,N+1)$ denotes the eigen energies of $H$, 
we have: $\Delta_{N+1}=\displaystyle{\prod_{n=0}^{N+1}(s+iE_n/\hbar)}$. As 
$\widehat{c}_{N+2}(s)=0$, we obtain from Eq.(\ref{rec}), 
\begin{equation}
\widehat{c}_{N+1}(s)=\frac{1}{\Delta_{N+1}}\,
\prod_{m=1}^{N+1}\left(-\frac{iv_m}{\hbar}\right)
=\frac{\displaystyle\prod_{m=1}^{N+1}\left(-\frac{iv_m}{\hbar}\right)}
{\displaystyle\prod_{n=0}^{N+1}(s+iE_n/\hbar)}\:.
\end{equation}
After inverse Laplace transforming this expression we find that the probability 
$P_{d\rightarrow a}(t)$ is given by Eq.(\ref{prob}) with the amplitude,   
\begin{eqnarray}
<\!N+1|\Pi_{n}|0\!>=\frac{\displaystyle{\prod_{i=1}^{N+1}v_i}}
{\displaystyle{
\underset{j\neq n}{\prod_{j=0}^{N+1}}(E_{n}-E_j)}}\:.
\end{eqnarray}
Now, using the formula in Eq.(\ref{hda2}) leads to the following expression 
for the effective coupling, 
\begin{equation}
|H_{da}|^2=\frac{\displaystyle{\prod_{i=1}^{N+1}v_i^2}}
{\displaystyle{\underset{j\neq l\neq l+1}{\prod_{j=0}^{N+1}}
(E_{l}-E_j)\,(E_{l+1}-E_j)}}\:,
\label{hda3}
\end{equation}
where the index ``$l$'' is defined such that $|\Delta E_{l+1,l}|$ is the 
smallest energy spacing of the eigen energy spectrum of the entire Hamiltonian. 
For this kind of Hamiltonian, the computation of the effective coupling is 
reduced to the determination of eigen energies of the system Hamiltonian. 

As a direct application of the formula in Eq.(\ref{hda3}), we consider 
the following illustrative examples in which the energies of the donor 
and acceptor are both equal, $\varepsilon_0=\varepsilon_{N+1}=\varepsilon$. 

{\bf $\bullet$ System with $2+1$ Levels:}

The simplest example which can worked out analytically is the one where the 
bridge is reduced to a single site (i.e., $N=1$). In this case, the Hamiltonian 
in Eq.(\ref{Ham}) reduces to a $3\,\times\,3$ matrix whose the eigen energies are 
given by, 
\begin{eqnarray}
\left\{\begin{array}{l}
E_0=\left[\varepsilon+\varepsilon_1-\sqrt{(\varepsilon-\varepsilon_1)^2+4
(v_1^2+v_2^2)}\right]/2\\
E_1=\varepsilon \\
E_2=\left[\varepsilon+\varepsilon_1+\sqrt{(\varepsilon-\varepsilon_1)^2+4
(v_1^2+v_2^2)}\right]/2
\end{array}\right.
\end{eqnarray}
It follows from this that the smallest energy spacing is $\Delta E_{1,0}$ with $l=0$ 
for $\varepsilon_1>\varepsilon$ and $\Delta E_{2,1}$ with $l=1$ for 
$\varepsilon_1<\varepsilon$. Applying the formula in Eq.(\ref{hda3}), we get:
\begin{eqnarray}
|H_{da}|^2=v_1^2v_2^2\times
\left\{\begin{array}{lcc}
1/\left[\Delta E_{0,2}\,\Delta E_{1,2}\right] & ; & \varepsilon_1>\varepsilon\\ \\
1/\left[\Delta E_{1,0}\,\Delta E_{2,0}\right] & ; & \varepsilon_1<\varepsilon
\end{array}\right.
\label{hdaN1}
\end{eqnarray}
This is to compare with $|H_{da}^{(1)}|=v_1v_2/|\varepsilon_1-\varepsilon|$, 
obtained from using the approximate formula in Eq.(\ref{hda1}). 

{\bf $\bullet$ Degenerate System with $2+2$ Levels:}

An other interesting example where the formula in Eq.(\ref{hda1}) fails to 
provide a finite value is, for instance, a bridge of size  $N=2$ with sites 
energies equal to the donor/acceptor energy, 
$\varepsilon_1=\varepsilon_2=\varepsilon$. In this case, the eigen energies of 
the $4\,\times\,4$ Hamiltonian matrix are given by, 
\begin{eqnarray}
\left\{\begin{array}{l}
E_0=\varepsilon-\lambda_+/2\\
E_1=\varepsilon-\lambda_-/2\\
E_2=\varepsilon+\lambda_-/2\\
E_3=\varepsilon+\lambda_+/2
\end{array}\right.
\end{eqnarray}
where $\lambda_{\pm}=\sqrt{2(v_1^2+v_2^2+v_3^2)\pm 
2\sqrt{(v_1^2+v_2^2+v_3^2)^2-4v_1^2v_3^2}}$, with $\lambda_+>\lambda_-$. Here, 
$l=0$ (or $l=2$) corresponding to the smallest energy spacing $\Delta E_{1,0}$  
(or $\Delta E_{3,2}$). As a result, we find: 
\begin{eqnarray}
|H_{da}|^2 & = & \frac{v_1^2v_2^2v_3^2}{\Delta E_{0,2}\,\Delta E_{1,2}\,
\Delta E_{0,3}\,\Delta E_{1,3}}\nonumber \\
& = & \frac{v_1v_2^2v_3}{4\left[(v_1+v_3)^2+v_2^2\right]}\:.
\end{eqnarray}

{\bf $\bullet$ System with $N+2$ Levels:}

As a last example, we consider the situation where the bridge of size $N$ is 
composed of sites with energies, $\varepsilon_n=\varepsilon_b$ and coupling, 
$v_n=v$. In this case, the determinant $\Delta_n$ for the eigen energies 
satisfies the recurrence relation, 
\begin{eqnarray}
\left\{\begin{array}{lcc}
\Delta_n-(\varepsilon_b-E)\Delta_{n-1}+v^2\Delta_{n-2}=0 & ; & n\leq N\\
\Delta_{N+1}-(\varepsilon-E)\Delta_{N}+v^2\Delta_{N-1}=0
\end{array}\right.
\end{eqnarray}
with $\Delta_{-1}=1$ and $\Delta_0=\varepsilon-E$. Introducing the variables, 
$\varepsilon_b-E=2vx$ and $\sigma=(\varepsilon_b-\varepsilon)/v$, one can 
show that,
\begin{eqnarray}
\left\{\begin{array}{lcc}
\Delta_n=v^{n+1}\,\left[U_{n+1}(x)-2\sigma\,U_{n}(x)\right] & ; & n\leq N\\
\Delta_{N+1}=2v(x-\sigma)\Delta_{N}-v^2\Delta_{N-1}=0
\end{array}\right.
\end{eqnarray}
where $U_{n}(\cos\theta)=\sin\left[(n+1)\theta\right]/\sin\theta$ is 
the trigonometrical representation for the Chebyshev polynomials of the 
second kind that satisfy the recurrence relation, 
$U_{n+1}(x)-2xU_{n}(x)+U_{n-1}(x)=0$ \cite{ABM}. The characteristic 
equation for the eigen energies reads, $\Delta_{N+1}=0$, i.e., 
\begin{equation}
U_{N+2}(x)-2\sigma\,U_{N+1}(x)+\sigma^2\,U_{N}(x)=0\:.
\label{char}
\end{equation}
As a polynomial of degree $N+2$ in $x$, this equation has $N+2$ roots 
$x_k^{(N)}(\sigma)$, $k=1,2,\cdots,N+2$, which could be found numerically. 
To each root corresponds an eigen energy, 
$E_{k-1}(\sigma)=\varepsilon_b-2vx_k^{(N)}(\sigma)$. One can convince ourselves 
when the eigen energies are arranged in ascending order (i.e., 
$E_0<E_1<\cdots<E_{N+1}$, a choice for the smallest energy spacing 
is $E_1-E_0$ (i.e., $x_1^{(N)}-x_2^{(N)}$) so that $l=0$. In this case, the 
effective coupling is given by, 
\begin{equation}
|H_{da}|^2=\frac{v^2}
{\displaystyle{4^{N}\prod_{j=2}^{N+1}\left[x_{1}^{(N)}-x_{j+1}^{(N)}\right]\times
\left[x_{2}^{(N)}-x_{j+1}^{(N)}\right]}}\:.  
\label{hda4}
\end{equation}
As a check, Eq.(\ref{char}) reduces for $N=1$ to, 
$8x^3-8\sigma x^2+2(\sigma^2-2)x+2\sigma=0$, whose solutions are: 
$x_1^{(1)}=\left[|\sigma|+\sqrt{\sigma^2+8}\right]/4$, 
$x_2^{(1)}=|\sigma|/2$ and 
$x_3^{(1)}=\left[|\sigma|-\sqrt{\sigma^2+8}\right]/4$. Using these roots in 
Eq.(\ref{hda4}) leads to Eq.(\ref{hdaN1}). 

The evaluation of $x_k^{(N)}(\sigma)$ in explicit terms does not seem to be feasible. 
However, for $\sigma=0$ or $\sigma=\pm 1$, Eq.(\ref{char}) has the exact solutions 
(for $k=1,2,\cdots,N+2$),
\begin{eqnarray}
x_k^{(N)}(0) & = & \cos\left[\frac{k\pi}{N+3}\right]\:, \\
x_k^{(N)}(\pm 1) & = & \cos\left[\frac{(k-1)\pi}{N+2}\right]\:. 
\end{eqnarray}
For comparison, the effective coupling obtained by using the approximate 
formula in Eq.(\ref{hda1}) is \cite{EK92,MKR94}, 
\begin{eqnarray}
|H_{da}^{(1)}|=\left|\frac{2^{N+1}\,v\sqrt{\sigma^2-4}}
{\left(\sigma+\sqrt{\sigma^2-4}\right)^{N+1}-
\left(\sigma-\sqrt{\sigma^2-4}\right)^{N+1}}\right|\:.
\label{hda5}
\end{eqnarray}
In contrast to Eq.(\ref{hda4}), this function diverges for $\sigma=0$ and values 
$\sigma_k^{(N)}=2\cos\left[k\pi/(N+1)\right]$, $k=1,2,\cdots,N$, corresponding to 
the donor/acceptor energy equal to eigen energies of the isolated bridge, 
$E_{k-1}^{\rm bridge}=\varepsilon_b-2\cos\left[k\pi/(N+1)\right]$. Numerical 
comparison of Eq.(\ref{hda5}) with the exact expression in Eq.(\ref{hda4}) shows 
that Eq.(\ref{hda5}) overestimates the value of the effective coupling. 

To sum up, we have derived a quite general and close formula in Eq.(\ref{hda2}) 
for computing the effective donor-acceptor coupling for a given Hamiltonian. 
For a tight-binding Hamiltonian with nearest-neighbor interactions, Eq.(\ref{hda2}) 
reduces to Eq.(\ref{hda3}) and the computation of the effective coupling is 
reduced to the determination of eigen energies of the system Hamiltonian. 
The presented results can be extended in several
directions to the more general processes of electron transfer between two reservoirs
of states, exemplified by the molecular vibrational levels associated with the donor and
acceptor sites connected by a molecular wire.
It requires a set of different methods with specific approximations on certain scales of length
and time but the coupling $H_{da}$ calculated in this paper is the indispensable entry point into such
more realistic study.

\end{document}